\documentclass{article}
\usepackage{PRIMEarxiv}

\usepackage{graphicx}%
\usepackage{multirow}%
\usepackage{amsmath,amssymb,amsfonts}%
\usepackage{amsthm}%
\usepackage{mathrsfs}%
\usepackage[title]{appendix}%
\usepackage{xcolor}%
\usepackage{textcomp}%
\usepackage{manyfoot}%
\usepackage{booktabs}%
\usepackage{algorithm}%
\usepackage{algorithmicx}%
\usepackage{algpseudocode}%
\usepackage{listings}%

\title{An accurate and robust level-set formulation for multiple junction kinetics} 

\author{Tianchi ~Li\thanks{Corresponding author: tianchi.li@minesparis.psl.eu} $^1$,
	Marc ~Bernacki$^1$ \\
	\\
	$^1$ Mines Paris, PSL University\\
 	Centre for material forming (CEMEF), UMR CNRS\\
 	 06904 Sophia Antipolis, France\\
}

\begin{document}
\maketitle

\begin{abstract}
The front-capturing level-set (LS) method is widely employed in academia and industry to model grain boundary (GB) migration during the microstructure evolution of polycrystalline materials under thermo-mechanical treatments. During capillarity-driven grain growth, the conventional mean curvature flow equation, $\vec{v} = - \mu \gamma \kappa \vec{n}$, is used to compute the GB normal migration velocity. Over recent decades, extensive efforts have been made to incorporate polycrystalline heterogeneity into this framework. However, despite increased complexity and computational costs, these approaches have yet to achieve fully satisfactory performance. This paper introduces a simple yet robust LS formulation that accurately captures multiple junction kinetics, even with extreme GB energy ratios. Validation against existing analytical solutions highlights the method’s accuracy and efficiency. This novel approach offers significant potential for advancing the study of highly heterogeneous interface systems.
\end{abstract}

\keywords{Microstructure Evolution, Grain Boundary Migration, Multiple Junction Kinetics, Front-Capturing, Level-Set, Finite Element.}

\maketitle

\section{Introduction}\label{sec1}

Most metallic materials used in modern industries exhibit polycrystalline microstructures, whose in-use properties are predominantly determined by the characteristics of these microstructures. Understanding and predicting the evolution of polycrystalline microstructures during various thermo-mechanical treatments in material forming processes are therefore crucial for optimizing material properties \cite{humphreys_2017}. With rapid advancements in computational infrastructure and methods, numerical modeling and simulation have become indispensable for studying microstructure evolution in polycrystalline materials \cite{chen_2014,bernacki2024digital}.

Full-field front-capturing approaches, such as the level-set (LS) and phase-field methods, are widely used to model microstructure evolution in polycrystals due to their ability to naturally incorporate complex morphological and topological changes during grain boundary (GB) migration \cite{osher_2001,chen_2002,bernacki2024digital}. For hot metal forming, where materials undergo large deformations, the LS method is often used in the state-of-the-art. Over recent decades, numerous LS frameworks, based on finite difference methods using regular grids, finite element approaches with structured or unstructured finite element meshes, and fast Fourier transformation techniques, have demonstrated their ability to statistically model the evolution of polycrystalline microstructures \cite{marc_2024}. However, existing models based on the well-known mean curvature flow equation, $\vec{v} = - \mu \gamma \kappa \vec{n}$, introduced by Mullins in 1956 \cite{mullins_1956}, sometimes struggle to interpret the complex kinetic behaviors of individual interfaces observed in recent experimental studies \cite{rohrer_2021,florez2022statistical}. This gap between experimental observations and numerical tools highlights the need for improved models that better capture the intricate kinetics of GBs, promoting relevant reverse engineering based on the comparison between simulation results and 3D \textit{in situ} data \cite{wang_2021}.

The kinetic properties of GBs, such as mobility $\mu$ and free energy $\gamma$, depend heavily on the underlying crystallography at the GBs. This leads to significant spatial heterogeneity of GB kinetic properties in polycrystalline materials \cite{smith_1948,Kohara_1958}. To address the effects of this heterogeneity on the migration of GBs and multiple junctions, i.e., fundamental components of polycrystalline microstructures, various continuum models have been proposed \cite{elsey_2013,miessen_2015,jin_2015,chang_2017}. Recently, a LS formulation was proposed to account for heterogeneous GB energies by recourse to the thermodynamics of grain growth \cite{julien_2018} and applied to simulate 2D grain growth of polycrystals \cite{julien_2020,brayan_2021}. Based on the disconnection-mediated GB migration theory, a continuum equation of motion was developed, initially for individual GBs and triple junctions (TJs) \cite{zhang_2017}, and later extended to the polycrystalline scale \cite{zhang_2021}. These recent advances aim to improve predictive accuracy by extending the conventional mean curvature flow mechanism with increasingly complex velocity expressions. Nevertheless, none of them fully captures the heterogeneity effects on the kinetic behavior of individual interfaces and multiple junctions.

In this work, we address the problem from a novel perspective. Rather than modifying the velocity definition, we introduce an additional source term on the right-hand side of the LS transport equation, while retaining the conventional diffusive term corresponding to the classical mean curvature flow mechanism. Validation against analytical solutions for quasi-static TJ migration velocity, equilibrium TJ dihedral angles and profiles demonstrates the accuracy and robustness of this formulation in capturing the effects of GB heterogeneity at multiple junctions, even for extreme GB energy ratios traditionally deemed unstable by existing methods. Furthermore, the proposed formulation is simplistic, computationally efficient, and straightforward to implement, avoiding the need for heavy computations of complex velocity terms.

\section{Results}\label{sec2}

\subsection{Level-Set formulation for grain boundary migration} \label{sec2-1}

Considering the physical space $\Omega$ of a polycrystalline material, its microstructure is represented by a network of grain boundaries (GBs) that divides $\Omega$ into a set of grains $G_i$, ($i=1,...,N$), where $N$ is the total number of grains. The level-set (LS) method describes each grain $G_i$ implicitly using a continuous scalar field $\psi_i$ initialized as the signed Euclidean distance to the grain's boundaries ($\Gamma_i = \partial G_i$) \cite{osher_1988}. The GBs correspond to the zero-isovalues of the LS functions:
\begin{equation}
    \begin{cases}
      \psi_i(x,t) = \pm d(x,\Gamma_i(t)), \; x \in \Omega \\
      \psi_i(x,t) = 0 \Leftrightarrow x \in \Gamma_i(t) .
    \end{cases} 
\end{equation}

\noindent
By convention, $\psi_i$ is assumed positive inside the grain $G_i$ and negative outside. This sign convention directly affects the definitions of the GB outward unit normal vector $\Vec{n}$ and the mean curvature $\kappa$, given by $\Vec{n} = - \Vec{\nabla}\psi / ||\Vec{\nabla} \psi||$, $\kappa = \Vec{\nabla} \cdot \Vec{n}$. The motion of GBs, characterized by the velocity field $\Vec{v}$, is governed by the temporal evolution of the LS functions through the transport equation: 
\begin{equation} \label{LS_transport}
    \frac{\partial \psi_i}{\partial t} + \Vec{v} \cdot \Vec{\nabla} \psi_i = 0 .
\end{equation}

\noindent
To preserve the distance metric property of the LS functions ($||\Vec{\nabla} \psi|| = 1$), a re-initialization step is often applied \cite{modesar_2015}. This simplifies the computation of $\Vec{n}$ and $\kappa$ to $\Vec{n} = - \Vec{\nabla} \psi $, $\kappa = - \Delta \psi$. For curvature-driven motion, substituting $\vec{v} = - \mu \gamma \kappa \vec{n}$ into the transport equation yields the LS formulation for mean curvature flow:
\begin{equation} \label{LS_transport_mcf}
    \frac{\partial \psi_i}{\partial t} - \mu \gamma \Delta \psi_i = 0 .
\end{equation}

\noindent
However, solving this equation alone can result in topological anomalies, such as overlaps or voids between adjacent grains. These kinematic incompatibilities are traditionally addressed using one of the following approaches: \\

1. Post-solution correction \cite{merriman_1994}: Resolve an additional equation to separate overlapping grains and close voids after solving Eq. \eqref{LS_transport_mcf}:
\begin{equation} \label{LS_correct_merriman}
    \psi_i(x,t) = \frac{1}{2} \left[ \psi_i(x,t) - \mathop{max}_{j \neq i} \psi_j(x,t) \right] ,
\end{equation} 

2. Source term penalization \cite{zhao_1996}: Include a source term in the LS transport equation to penalize the appearance of overlaps and voids:
\begin{equation} \label{LS_correct_zhao}
    \frac{\partial \psi_i}{\partial t} - \mu\gamma\Delta\psi_i = \lambda\mu \left( 1 - \sum_{j=1}^{n} H(\psi_j) \right) ,
\end{equation}

\noindent
where $H$ is the Heaviside function and $\lambda$ is a Lagrange multiplier associated with a minimization problem. 

While effective, these strategies fail to respect GB heterogeneity. For instance, in 2D, they enforce equal dihedral angles near triple junctions, which is only valid for uniform GB properties. This limitation hinders the accurate representation of GB heterogeneity, especially at junctions where GBs with different properties meet.

To address this limitation, we propose a novel LS formulation:
\begin{equation} \label{LS_transport_Tianchi}
    \frac{\partial \psi_i}{\partial t} - \mu\gamma\Delta\psi_i = \Lambda_i\mu \left( 1 - \sum_{j=1}^{n} H(\psi_j) \right) .
\end{equation}

\noindent
Although similar in form to Eq. \eqref{LS_correct_zhao}, this formulation fundamentally differs in its treatment of heterogeneity. Each LS function $\psi_i$, representing a grain, is associated with an auxiliary scalar field $\Lambda_i$, which is defined based on the kinetic properties of the grain's boundaries. Since $\Lambda_i$ is defined in the same space as $\psi_i$, this approach is fully compatible with existing numerical strategies at the polycrystalline scale, including the regrouping of non-adjacent grains into global LS functions using coloring techniques to enhance computational efficiency in large-scale simulations \cite{krill_2002,elsey_2009,scholtes_2015}. 

Unlike other LS formulations that modify $\mu$ and $\gamma$ or introduce additional convective terms to account for GB heterogeneity \cite{hallberg_2019,brayan_2021}, our approach preserves the conventional mean curvature flow description. Heterogeneous GB properties are independently incorporated through a source term on the right-hand side of the LS transport equation. This formulation provides a unified framework that seamlessly integrates the treatment of GB heterogeneity with the correction of kinematic incompatibilities. In the following, this model is developed and validated in the context of a disorientation angle-like dependency of $\gamma_i$, and consequently, $\Lambda_i$.

\subsection{Validation against analytical solutions} \label{sec2-2}

\begin{figure}[!ht]
    \centering
    \includegraphics[width=0.75\textwidth]{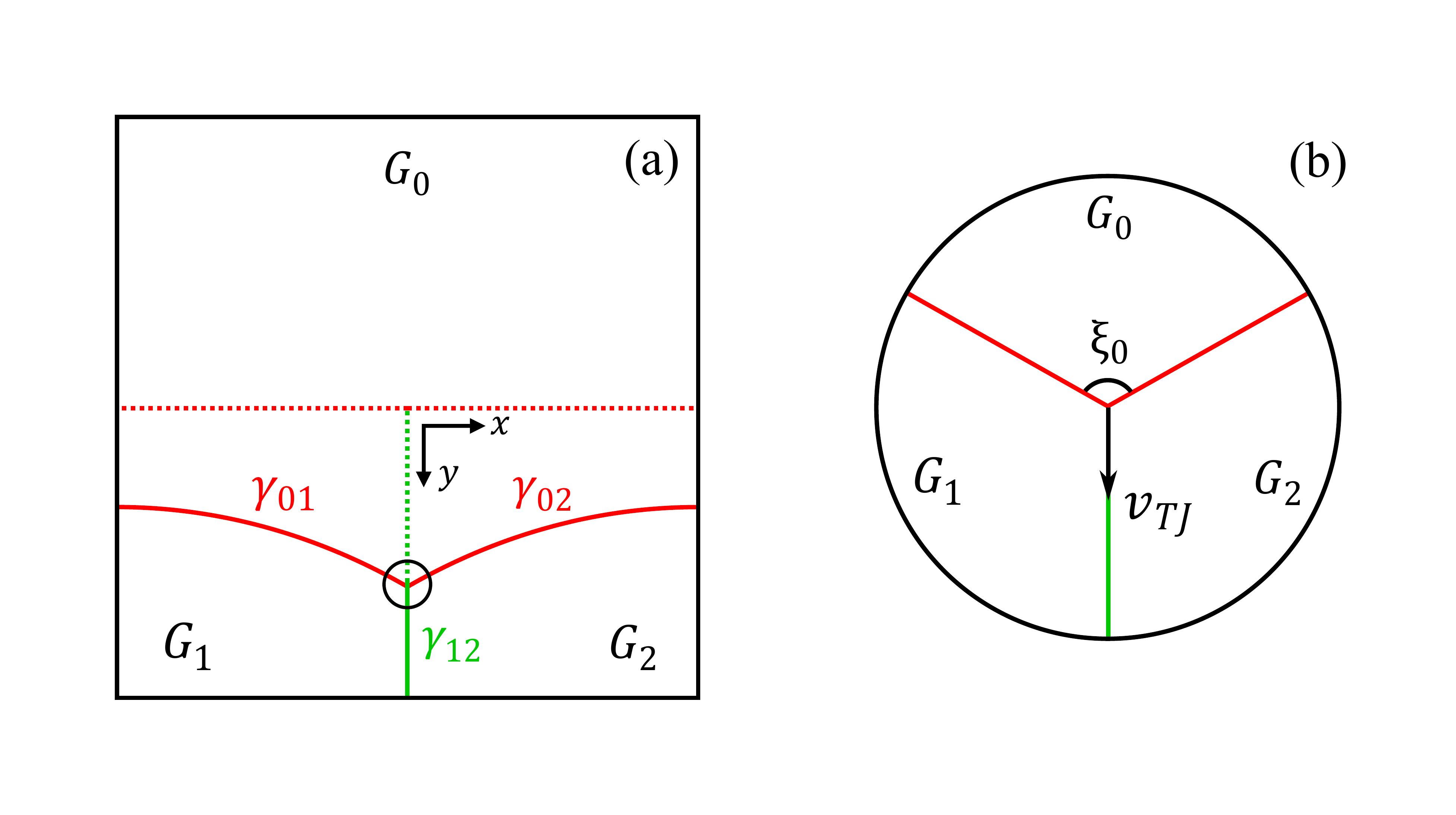}
    \caption{Schematic illustration for \textbf{(a)} Garcke's analytical case with \textbf{(b)} zoom on TJ point.}
    \label{Garcke}
\end{figure}

To evaluate the performance of this novel level-set (LS) formulation in capturing the effects of grain boundary (GB) heterogeneity on the microstructure evolution of polycrystals, it is validated thanks to the well-known 2D analytical case proposed by Garcke \cite{garcke_1999}. As illustrated in Fig.\ref{Garcke}, this case describes the capillarity-driven migration of an initially T-shaped symmetric triple junction (TJ). The curvature-driven evolution of this representative three-grain system consists of two stages: an initial transient state followed by a quasi-static state. In the quasi-static state, analytical solutions have been established for \\

\noindent 1) the top dihedral angle at the TJ:
\begin{equation} \label{Garcke_Angle}
    \xi_0 = 2 \arccos{\left(\frac{1}{2R_\gamma}\right)} ,
\end{equation}

\noindent 2) the TJ migration velocity:
\begin{equation} \label{Garcke_Velocity}
    v_{TJ} = \frac{\mu\gamma_{top}}{L_x}\left(\pi-\xi_0\right) ,
\end{equation}

\noindent 3) the TJ profile:
\begin{equation} \label{Garcke_Profile}
    y(x,t)=v_{TJ}t + \frac{\mu\gamma_{top}}{v_{TJ}} \ln{\left(\cos{\left(\frac{v_{TJ}}{\mu\gamma_{top}}\left(\frac{L_x}{2}-\left|x\right| \right)\right)}\right)} , \qquad x \in \left[ -\frac{L_x}{2}, \frac{L_x}{2} \right]
\end{equation}

\noindent where $R_\gamma$ is the free energy ratio between the top and bottom GBs ($R_\gamma = \gamma_{top}/\gamma_{bot}$), and $L_x$ is the width of the simulation domain. In the axially symmetric configuration, $\gamma_{top} = \gamma_{01} = \gamma_{02}$, $\gamma_{bot} = \gamma_{12}$ (see Fig. \ref{Garcke}). Using a dimensionless setting ($\mu = \gamma_{top} = L_x = 1$), the analytical expressions for TJ velocity and profile simplify to
\begin{equation} \label{Garcke_Velocity_dimensionless}
     v_{TJ} = \pi-\xi_0 , \qquad \xi_0 \in \left(0, \pi\right],
\end{equation}

\begin{equation} \label{Garcke_Profile_dimensionless}
     y(x,t) = v_{TJ}t + \frac{1}{v_{TJ}} \ln{\left(\cos{\left(v_{TJ} \left(\frac{1}{2}-\left|x\right| \right)\right)}\right)}, \qquad x \in \left[ -\frac{1}{2}, \frac{1}{2} \right].
\end{equation}
The above analytical solutions are valid for the entire heterogeneity range beyond wetting limit, i.e., $R_\gamma > 1/2$.

\begin{figure}[!ht]
    \vspace{0.5 cm}
    \centering
    \includegraphics[width=0.75\textwidth]{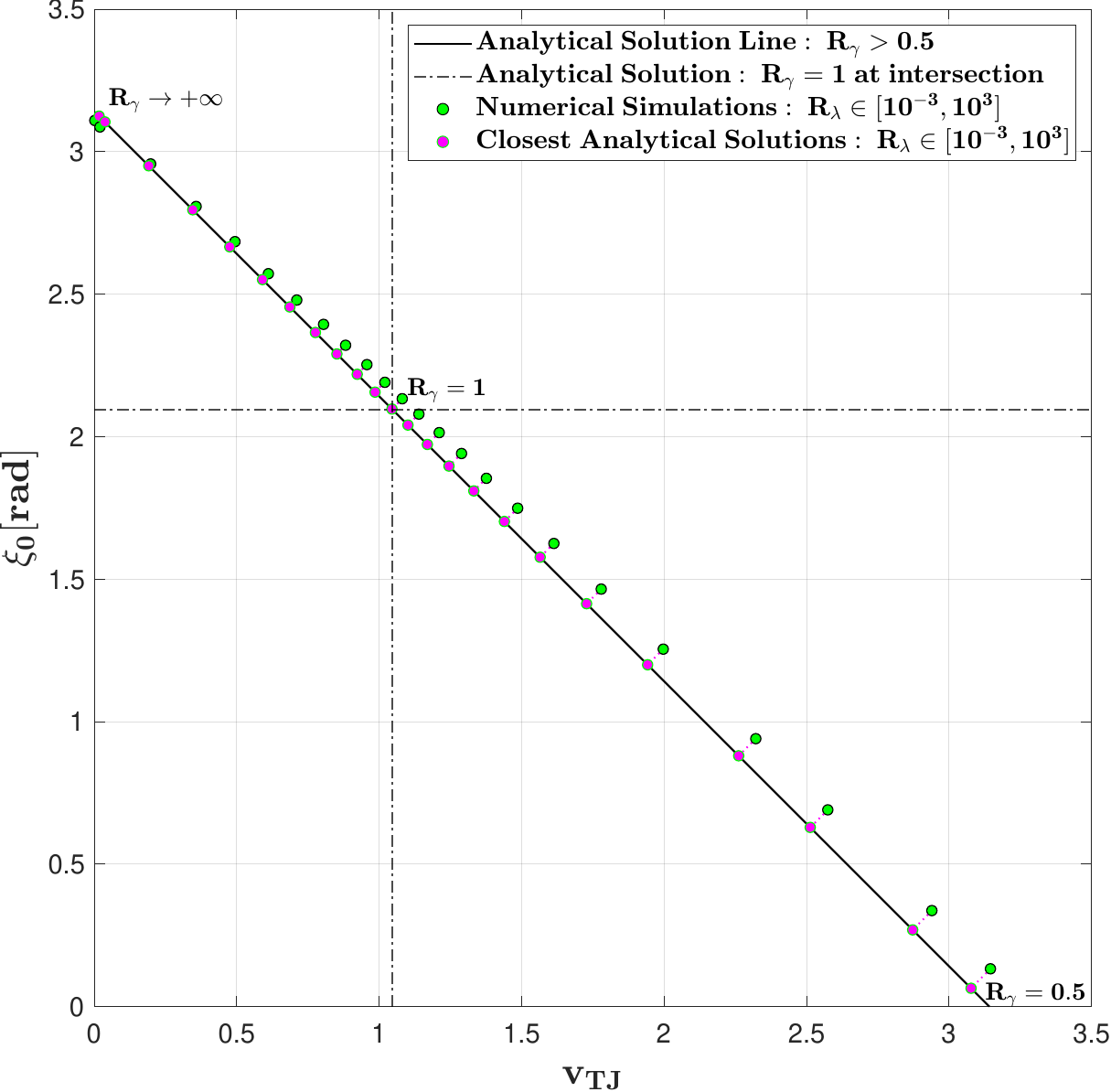}
    \caption{Numerically obtained quasi-static TJ top dihedral angle $\xi_0$ and migration velocity $v_{TJ}$ along with their projections on the analytical solution line: $v_{TJ} = \pi-\xi_0$. }
    \label{Angle_Velocity}
\end{figure}

\noindent
The dimensionless analytical case of Garcke is particularly suited for validating GB kinetic models in heterogeneous settings, since the exact solutions depend exclusively on the GB energy ratio $R_\gamma$, which reflects the inherent heterogeneity of the polycrystalline system. In line with the analytical configuration, where the GB energy is constant per interface and axially symmetric, we simplify here the $\Lambda_i$ field as a constant per grain and also axially symmetric ($\Lambda_{top}=\Lambda_0$, $\Lambda_{bot}=\Lambda_1=\Lambda_2$). Numerical tests reveal that the quasi-static kinetics of TJs depends only on the ratio of $\Lambda_i$ values rather than their absolute magnitudes. Further simplification is hence achieved by normalizing $\Lambda_i$ through division by their maximum value, $\max(\Lambda_i)$, yielding $\lambda_i = \Lambda_i / \max(\Lambda_i)$, with $\lambda_i \in (0,1]$. The ratio $R_\lambda=\lambda_{top}/\lambda_{bot}$ is then analogous to $R_\gamma$. Since $\max(\Lambda_i)$ is independent of the kinetic problem, its value is chosen to minimize the remaining vacuum at the TJ. For all simulations presented in this work, we consider a constant value of 600 for $\max(\Lambda_i)$.

To demonstrate the accuracy and robustness of this LS formulation in replicating TJ kinetics, we conducted numerical simulations using Garcke's dimensionless configuration for a set of $\{\lambda_{top},\lambda_{bot}\}$ (see table \ref{table_lambda}), covering the range $R_\lambda \in [10^{-3}, 10^3]$. 

\begin{table}[h]
\caption{The set of $\{\lambda_{top}$, $\lambda_{bot}\}$ used in numerical simulations}
\label{table_lambda}
\begin{tabular*}{\textwidth}{@{\extracolsep\fill}ccccccccccccc}
\toprule%
$\lambda_{top}$ & $10^{-3}$ & $10^{-2}$ & 0.1 & 0.2 & 0.3 & 0.4 & 0.5 & 0.6 & 0.7 & 0.8 & 0.9 & 1  \\
$\lambda_{bot}$ & 1 & 1 & 1 & 1 & 1 & 1 & 1 & 1 & 1 & 1 & 1 & 1 \\
\midrule
$\lambda_{top}$ & 1 & 1 & 1 & 1 & 1 & 1 & 1 & 1 & 1 & 1 & 1 & 1 \\
$\lambda_{bot}$ & 0.9 & 0.8 & 0.7 & 0.6 & 0.5 & 0.4 & 0.3 & 0.2 & 0.1 & 0.05 & $10^{-2}$ & $10^{-3}$  \\
\bottomrule
\end{tabular*}
\end{table}

\begin{figure}[!ht]
    \centering
    \includegraphics[width=\textwidth]{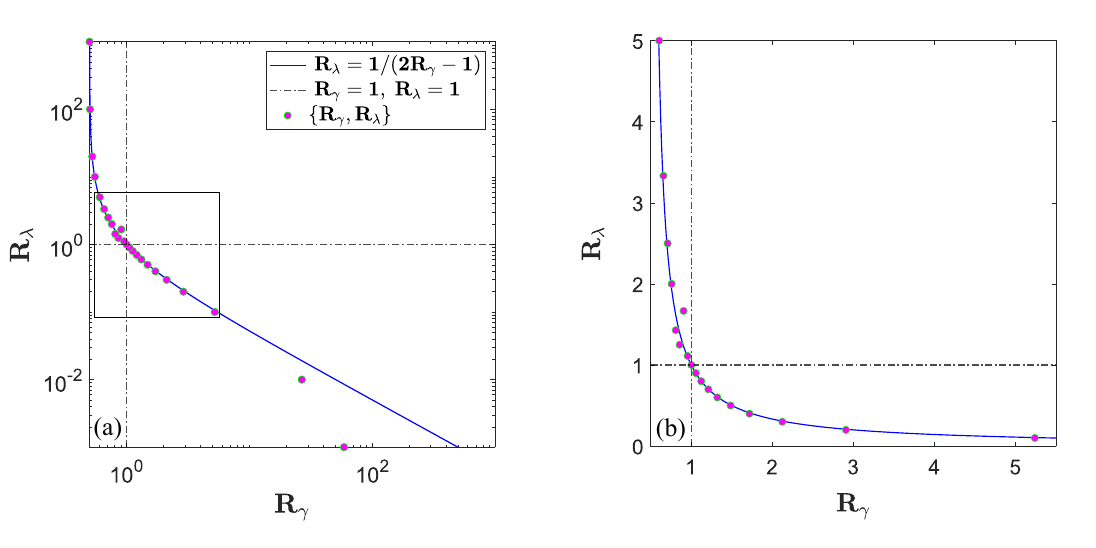}
    \caption{\textbf{(a)} Relation between $R_\gamma$ and $R_\lambda$ with \textbf{(b)} zoom on low to medium heterogeneity range.}
    \label{R_gamma_R_lambda}
\end{figure}

In Fig. \ref{Angle_Velocity}, the quasi-static values of $\{\xi_0,v_{TJ}\}$ obtained from simulations using $\{\lambda_{top},\lambda_{bot}\}$ as specified in Table \ref{table_lambda} are presented. The formulation accurately reproduces the TJ kinetics across the entire heterogeneous domain ($R_\gamma > 0.5$). By projecting the numerical solutions onto the analytical solution line, one can extract the corresponding analytical values of $\{\xi_0,v_{TJ}\}$ and calculate $R_\gamma$ using Eq. \eqref{Garcke_Angle}. Plotting these values of $R_\gamma$ against $R_\lambda$ (Fig. \ref{R_gamma_R_lambda}), one obtains the following apparent relationship between $\gamma_i$ and $\lambda_i$ for a large range of $R_\gamma$ values:

\begin{equation} \label{Relation_R_gamma_R_lambda}
    R_\lambda = \frac{1}{2R_\gamma-1} .
\end{equation}

\begin{figure}[!ht]
    \centering
    \includegraphics[width=0.75\textwidth]{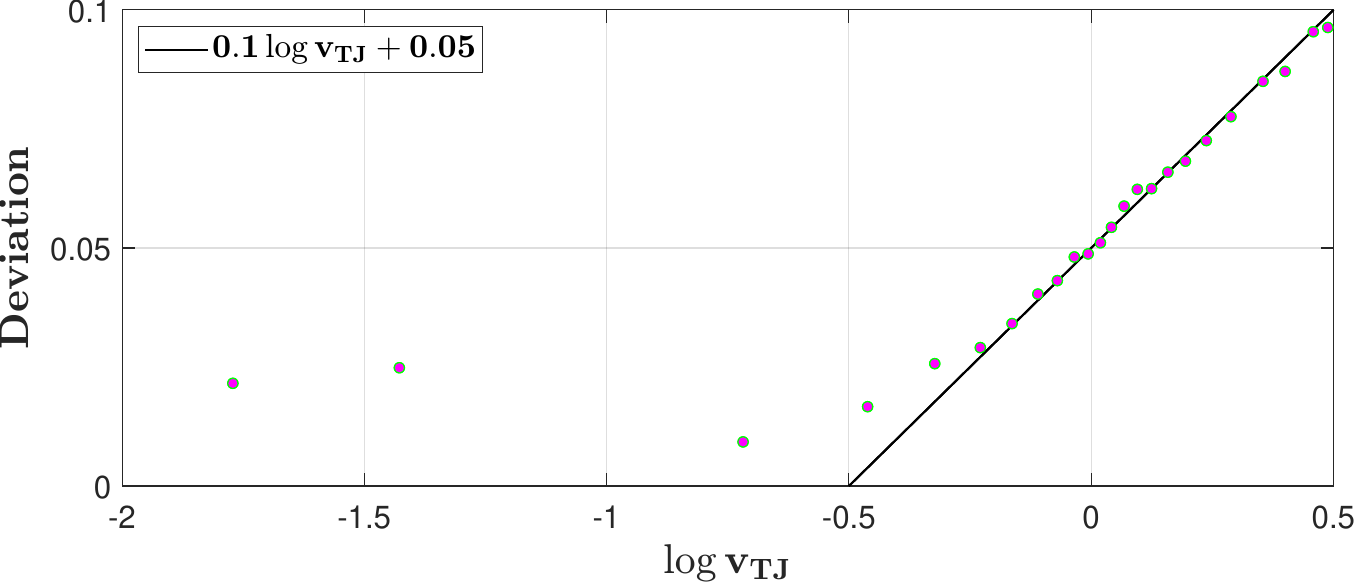}
    \caption{Deviation of the numerical solutions $\{\xi_0,v_{TJ}\}$ from the analytical solution line $v_{TJ} = \pi-\xi_0$.}
    \label{Error_Velocity}
\end{figure}

\begin{figure}[!ht]
    \vspace{1 cm}
    \centering
    \includegraphics[width=0.75\textwidth]{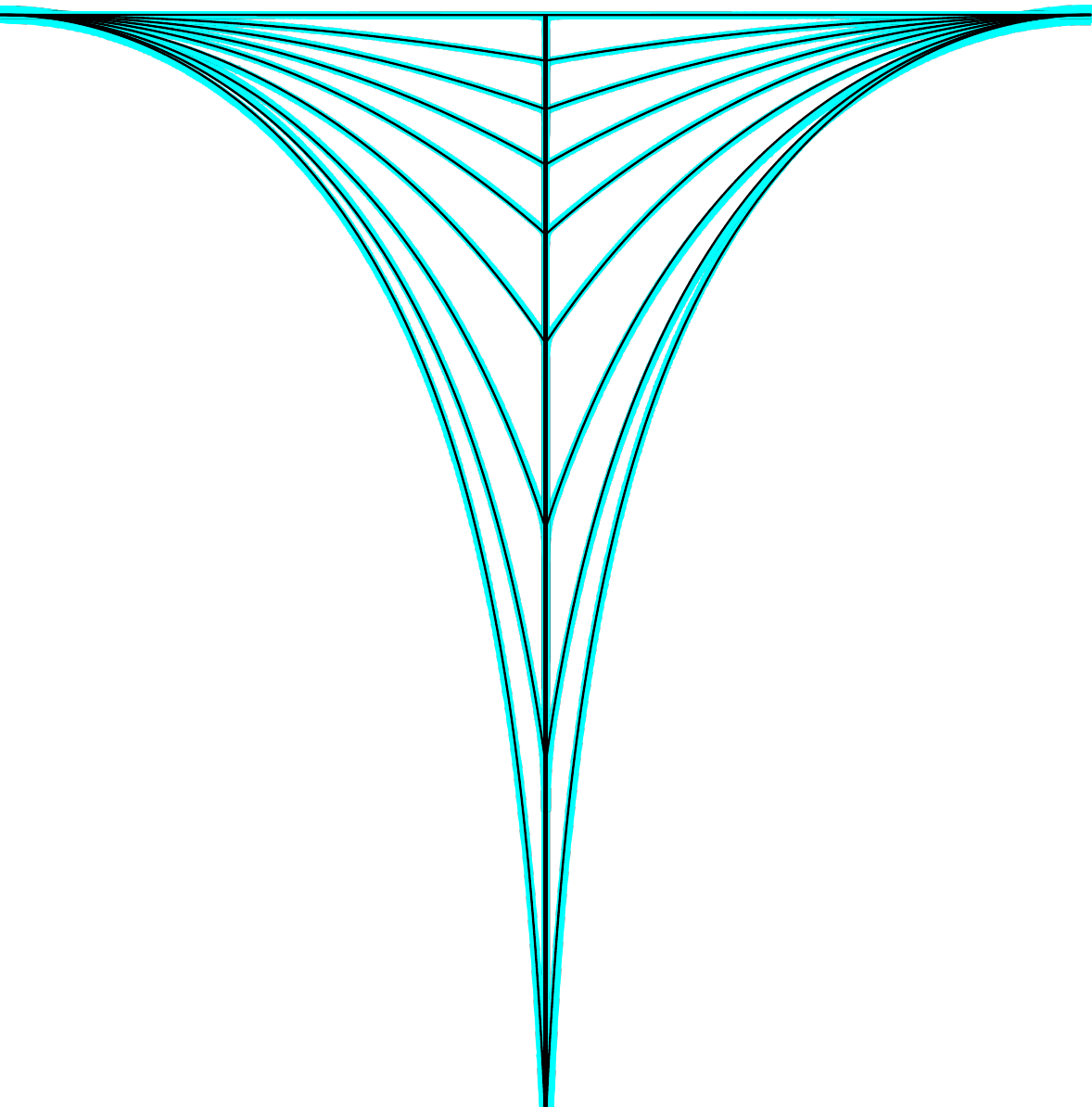} 
    \vspace{20 pt}
    \caption{Comparison between 2D \textbf{(Cyan)} numerical and \textbf{(Black)} analytical TJ profiles in quasi-static regimes (see Eq. \eqref{Garcke_Profile_dimensionless}).} 
    \label{Garcke_Profile}
\end{figure}

In the dimensionless configuration, the deviation of the numerical solutions from the analytical predictions is estimated as the distance between the numerically obtained data points $\{\xi_0, v_{TJ}\}$ and the analytical solution line $v_{TJ} = \pi - \xi_0$. As shown in Fig. \ref{Error_Velocity}, this deviation increases logarithmically with $v_{TJ}$ in the high-velocity regime, reaching a maximum of approximately 0.1. A sensitivity analysis confirms that our numerical solutions converge in both space and time to the analytical predictions.

In Fig. \ref{Garcke_Profile}, we present quasi-static TJ profiles obtained from simulations using a selected set of $R_\lambda$ values listed in Table \ref{table_R_lambda_R_gamma_profile}. The corresponding analytical TJ profiles are derived by first computing the $R_\gamma$ values with Eq. \eqref{Relation_R_gamma_R_lambda} and then using successively Eqs. \eqref{Garcke_Angle} \eqref{Garcke_Velocity_dimensionless} \eqref{Garcke_Profile_dimensionless}. As shown in Fig. \ref{Garcke_Profile}, while vertically aligned at the top, the numerical TJ profiles exhibit excellent agreement with their analytical counterparts.

\begin{table}[h]
\caption{The set of $\{R_\lambda$, $R_\gamma\}$ used to generate the numerical and analytical TJ profiles in Fig. \ref{Garcke_Profile}}
\label{table_R_lambda_R_gamma_profile}
\begin{tabular*}{\textwidth}{@{\extracolsep\fill}cccccccccc}
\toprule%
$R_\lambda$ & $10^{-3}$ & 0.2 & 0.5 & 1 & 2 & 5 & 20 & $10^{2}$ & $10^{3}$ \\
$R_\gamma$ & 500.5 & 3 & 1.5 & 1 & 0.75 & 0.6 & 0.525 & 0.505 & 0.5005 \\
\bottomrule
\end{tabular*}
\footnotetext{Note: $R_\gamma$ are obtained by substituting $R_\lambda$ into Eq. \eqref{Relation_R_gamma_R_lambda}.}
\end{table}

\begin{figure}[!ht]
    \vspace{0.5 cm}
    \centering
    \includegraphics[width=0.9\textwidth]{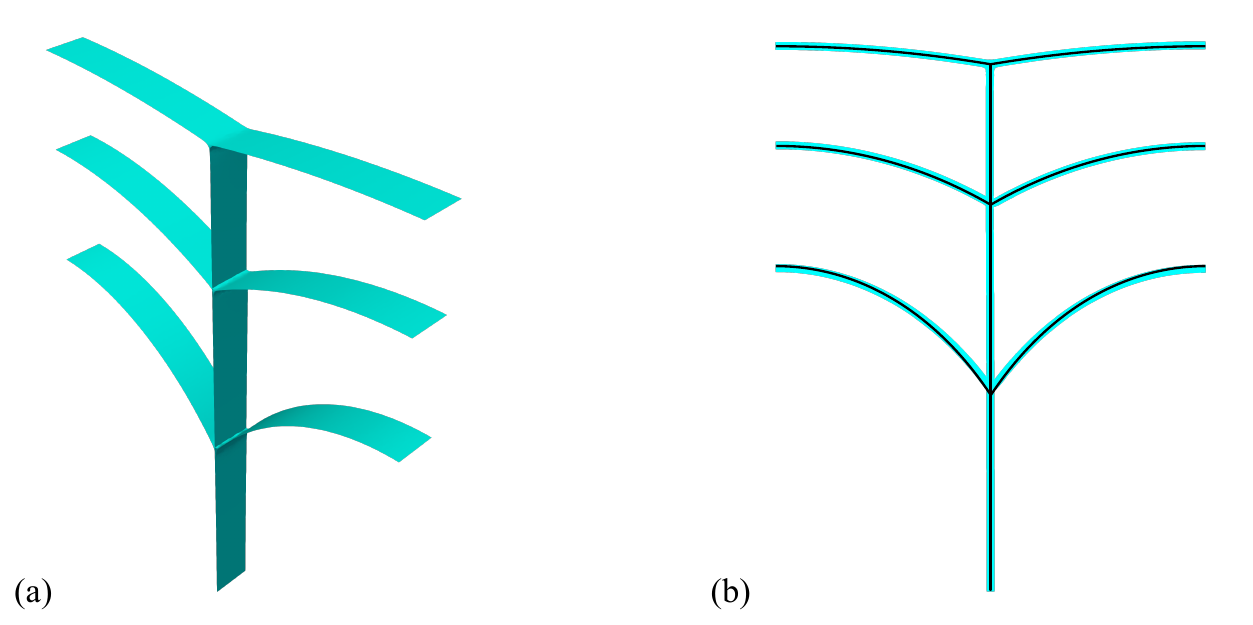} 
    \caption{\textbf{(a)} Equilibrium 3D TJ profiles simulated with $R_\lambda = 0.2$, $1$, $5$. \textbf{(b)} Comparison with 2D analytical profiles for $R_\gamma = 3$, $1$, $0.6$ (see Table \ref{table_R_lambda_R_gamma_profile}) depicted in black.} 
    \label{TJ_3D}
\end{figure}

To extend our validation to 3D, we extrude the 2D TJ point into a 3D TJ line of length 0.1 and rerun the simulations for $R_\lambda = 0.2$, $1$, $5$. When projected along the TJ line, the 3D numerical TJ profiles again closely match the analytical predictions (see Fig. \ref{TJ_3D}).  

With this, the proposed LS approach is validated against all analytical predictions proposed by Garcke across the entire heterogeneity range. This LS front-capturing formulation achieves unprecedented accuracy and robustness in modeling multiple junctions, even in the presence of extreme interface heterogeneity generally neglected in the discussion of existing front-capturing models \cite{hoffrogge_2025,jin_2015}.

\subsection{Simulation of an arbitrary triple junction} \label{sec2-3}

In real-world applications, polycrystalline microstructures often exhibit complex grain boundary (GB) heterogeneities, leading to arbitrary triple junction (TJ) configurations. Therefore, it is essential to validate our novel level-set (LS) formulation in simulating the migration of arbitrary TJs, ensuring its applicability to realistic scenarios. However, for arbitrary TJs, analytical solutions are only available for the TJ dihedral angles, which can be determined using Herring’s equilibrium \cite{herring_1999}. In our case, where GB energy remains constant for each interface, this equilibrium condition simplifies to Young’s equation:

\begin{equation} \label{Equation_Young} 
\frac{\sin{\xi_0}}{\gamma_{12}} = \frac{\sin{\xi_1}}{\gamma_{02}} =\frac{\sin{\xi_2}}{\gamma_{01}}, 
\end{equation} 

\noindent where $\xi_i$ denotes the dihedral angle within grain $G_i$, and $\gamma_{ij}$ represents the free energy of the GB separating grains $G_i$ and $G_j$, for all $i,j \in {1,2,3}$ and $i \neq j$.

\begin{figure}[!ht] 
\centering 
\includegraphics[width=\textwidth]{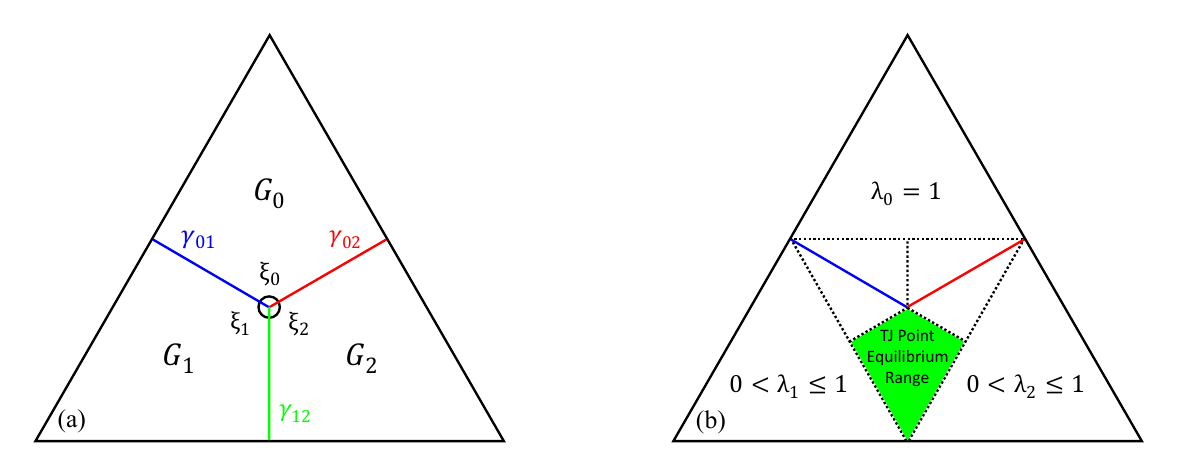} \caption{Schematic representation of \textbf{(a)} the dimensionless triangular test case and \textbf{(b)} the range of $\lambda_i$ values along with the corresponding range of equilibrium TJ positions.} \label{Young} 
\end{figure}

\begin{figure}[!ht] 
\vspace{0.5 cm}
\centering 
\includegraphics[width=0.85\textwidth]{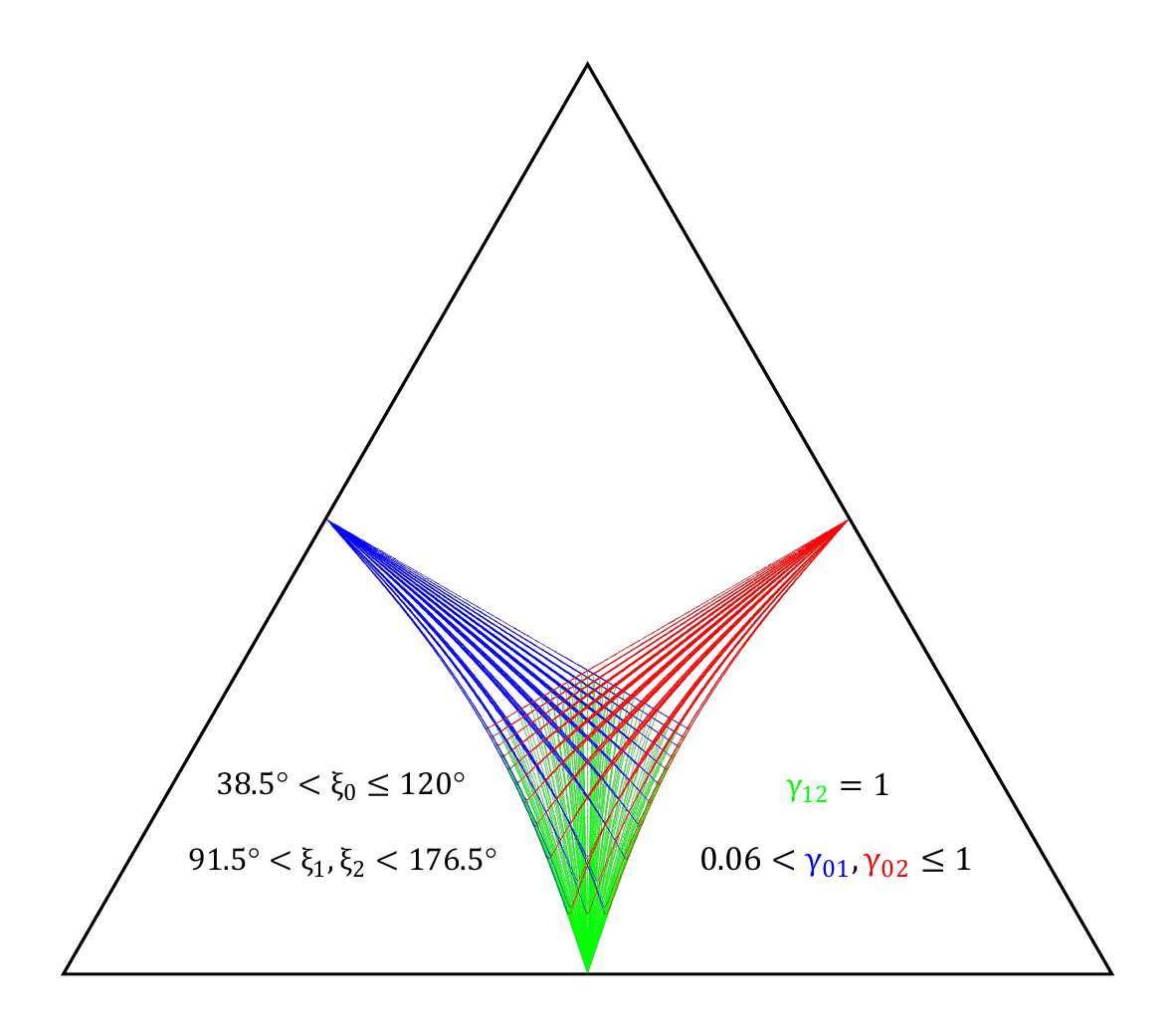} \caption{161 equilibrium TJ profiles simulated using the dimensionless triangular test case and the corresponding range of $\xi_i$ and $\gamma_{ij}$.} \label{Young_Profile} 
\end{figure}

\begin{figure}[!ht]
\centering
\includegraphics[width=0.8\textwidth]{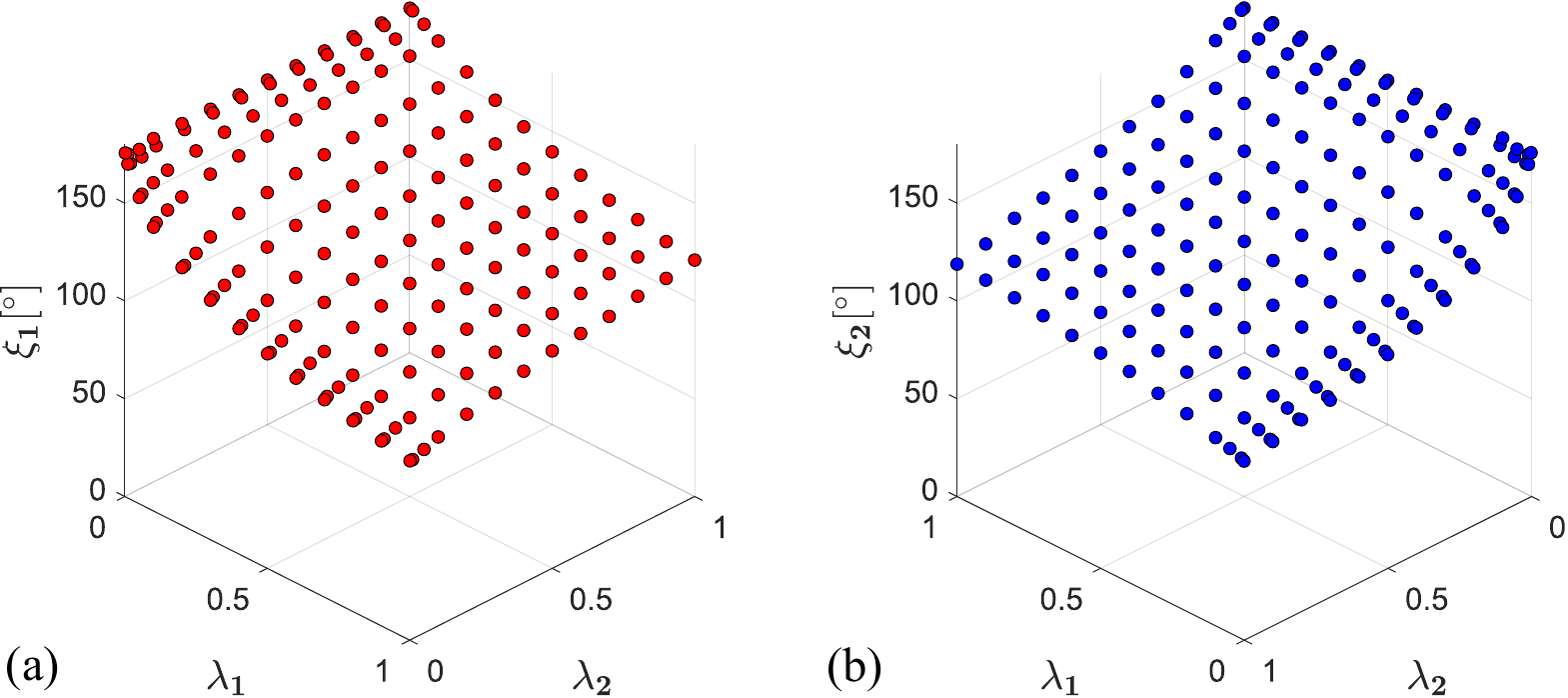}
\caption{169 measured values of \textbf{(a)} $\xi_1$ and \textbf{(b)} $\xi_2$ as functions of $\lambda_1$ and $\lambda_2$.}
\label{Young_Lambda_Angle}
\end{figure}

\begin{figure}[!ht]
\vspace{0.25 cm}
\centering
\includegraphics[width=0.8\textwidth]{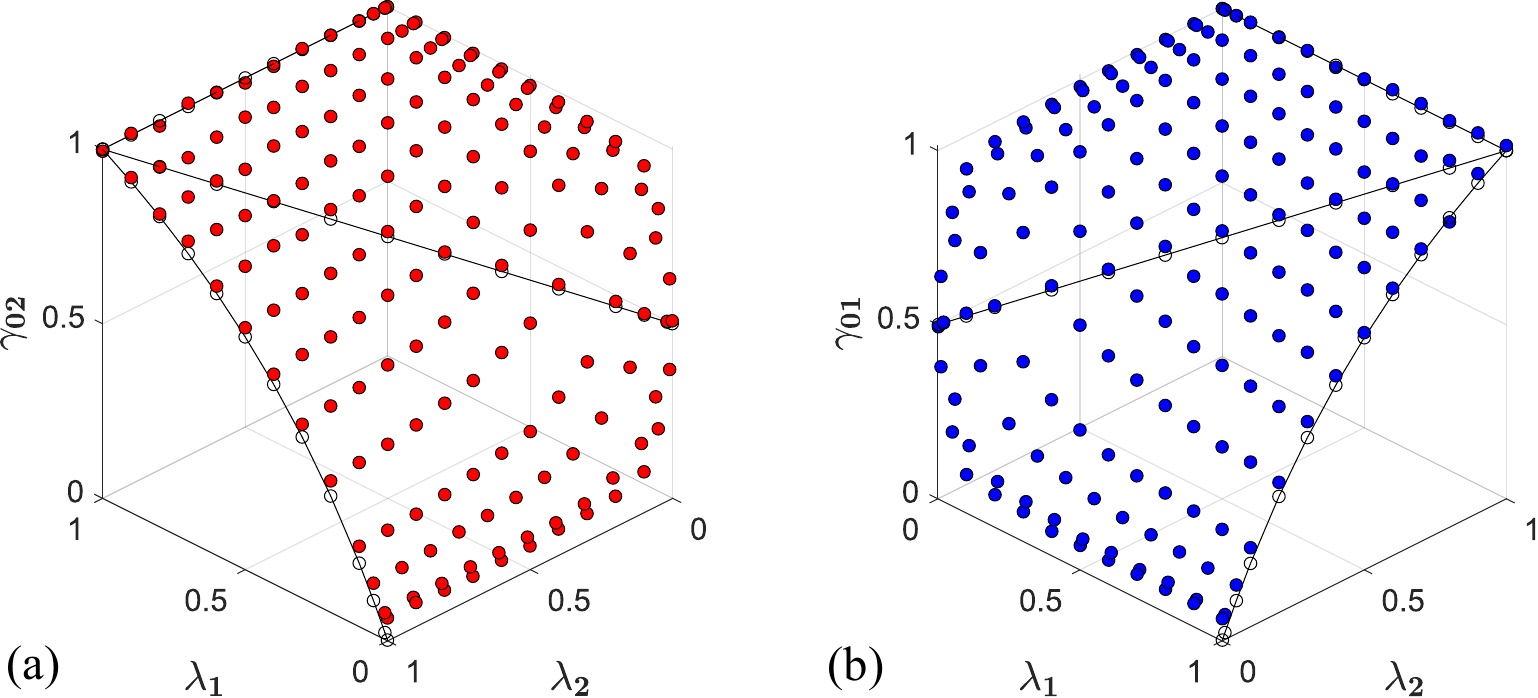}
\caption{\textbf{(a)} $\gamma_{02}$ and \textbf{(b)} $\gamma_{01}$ as functions of $\lambda_1$ and $\lambda_2$. For symmetric configurations, the predictions given by Eq. \eqref{Relation_R_gamma_R_lambda} are depicted in black.}
\label{Young_Lambda_Gamma}
\end{figure}

\begin{figure}[!ht]
\vspace{0.25 cm}
\centering
\includegraphics[width=0.8\textwidth]{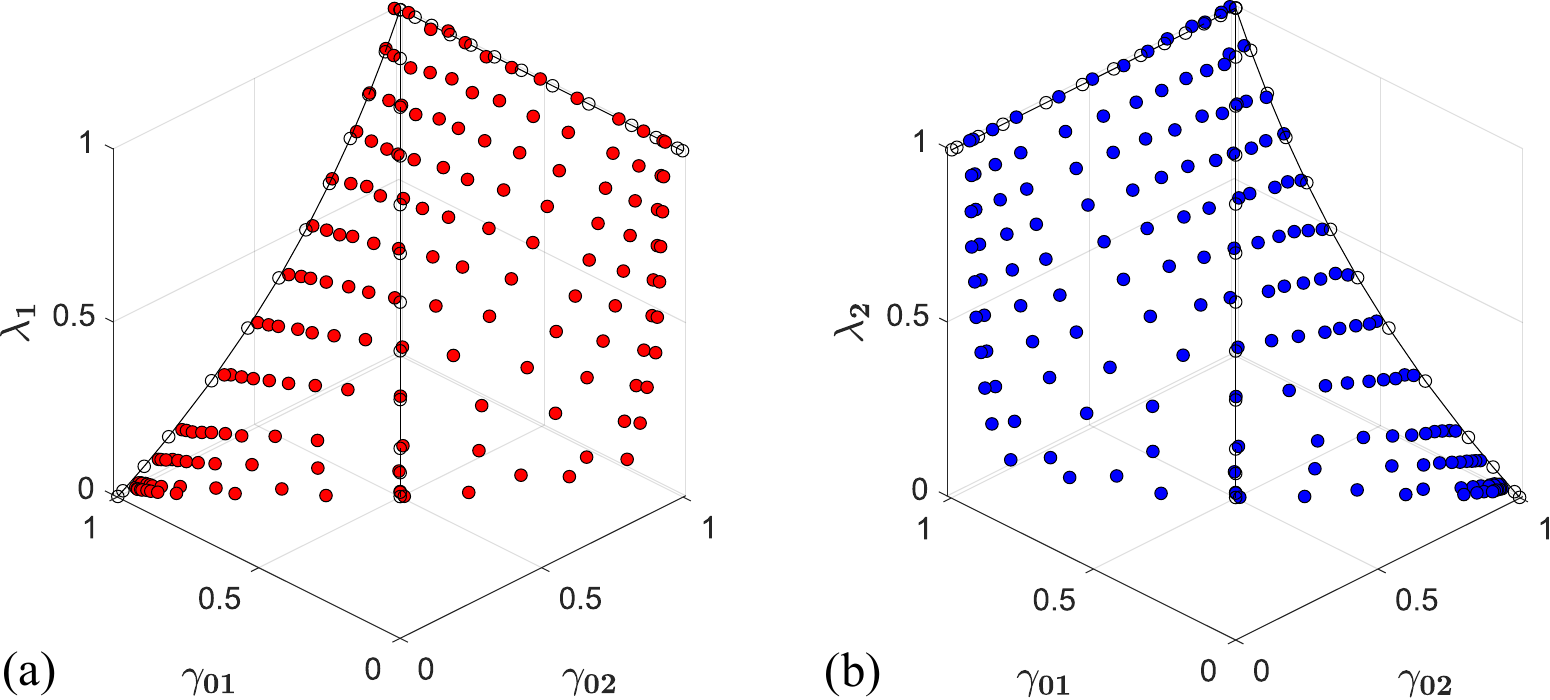}
\caption{\textbf{(a)} $\lambda_1$ and \textbf{(b)} $\lambda_2$ as functions of $\gamma_{02}$ and $\gamma_{01}$. For symmetric configurations, the predictions given by Eq. \eqref{Relation_R_gamma_R_lambda} are depicted in black.}
\label{Young_Gamma_Lambda}
\end{figure}

To validate our LS formulation against Young's equation for arbitrary triple junction (TJ) migration, we adopt the dimensionless triangular test case introduced in \cite{julien_2018}. The initial configuration is axially symmetric, with the TJ positioned at the incenter of an equilateral triangular domain, where the incircle diameter is 1. The three grain boundaries (GBs) connect to the midpoints of the triangle’s sides, each with a length of 0.5, forming equal dihedral angles of $120^\circ$ (see Fig. \ref{Young}(a)). Exploiting this symmetry, we reduce the domain of interest to one-third of the triangle while still capturing all possible TJ profiles. Dirichlet boundary conditions are applied to fix the intersection points of the three GBs with the domain boundaries, thereby limiting the range of TJ migration, as illustrated in Fig. \ref{Young}(b).  

For the numerical simulations, we fix $\lambda_0 = 1$ and vary $\lambda_1$ and $\lambda_2$ between 0 and 1. Specifically, they take 13 values: \{$10^{-3}$, $10^{-2}$, 0.05, 0.1, 0.2, 0.3, 0.4, 0.5, 0.6, 0.7, 0.8, 0.9, 1\}, resulting in a total of $13 \times 13 = 169$ parameter combinations. Fig. \ref{Young_Profile} displays the equilibrium TJ profiles for 161 of these combinations, excluding the 8 cases where $\lambda_1 + \lambda_2 < 0.1$, as the imposed boundary conditions strongly influence the TJ equilibrium in these cases. The measured equilibrium dihedral angles $\xi_1$ and $\xi_2$ for all 169 cases are shown in Fig. \ref{Young_Lambda_Angle} as functions of $\lambda_1$ and $\lambda_2$. For $\lambda_1 + \lambda_2 < 0.1$, the results are obtained separately using the benchmark configuration presented in the previous subsection.  
Under the dimensionless assumption $\gamma_{12} = 1$, the corresponding values of $\gamma_{02}$ and $\gamma_{01}$ are determined using Young’s equation: $\gamma_{02} = \sin{\xi_1}/\sin{\xi_0}$, $\gamma_{01} = \sin{\xi_2}/\sin{\xi_0}$, where $\xi_0 = 2\pi - \xi_1 - \xi_2$. The resulting values of $\gamma_{02}$ and $\gamma_{01}$ are visualized as functions of $\lambda_1$ and $\lambda_2$ in Fig. \ref{Young_Lambda_Gamma}. Conversely, $\lambda_1$ and $\lambda_2$ are plotted as functions of $\gamma_{02}$ and $\gamma_{01}$ in Fig. \ref{Young_Gamma_Lambda}. For three symmetric configurations: (1) $\lambda_1=\lambda_2$, (2) $\lambda_0=\lambda_2$, and (3) $\lambda_0=\lambda_1$, we also include the predictions given by Eq. \eqref{Relation_R_gamma_R_lambda}. 

As demonstrated in Figs. \ref{Young_Lambda_Angle}, \ref{Young_Lambda_Gamma}, and \ref{Young_Gamma_Lambda}, any dihedral angle configuration $\{\xi_1,\xi_2\}$ within the realistic heterogeneous domain (beyond the wetting limit, i.e., $\gamma_{02} + \gamma_{01} > 1$) can be accurately reproduced by the LS formulation using an appropriate set of $\{\lambda_1, \lambda_2\}$ corresponding to $\{\gamma_{02}, \gamma_{01}\}$ from the physical problem. For practical applications, one may use numerically obtained data points to interpolate the relationship between $\gamma$ and $\lambda$. The proposed LS approach exhibits strong potential for fully capturing the evolution of real-world heterogeneous interface systems.

\section{Discussion}\label{sec3}

In this paper, a novel level-set (LS) formulation is presented for simulating curvature-driven migration of multiple junctions, a fundamental structural feature in materials with interface network microstructures, such as polycrystals. This formulation is simple in form, straightforward to implement, and computationally efficient compared to existing approaches. Notably, it achieves unprecedented accuracy and robustness in capturing the effects of heterogeneity on multiple junction kinetics by decoupling the heterogeneous term from the conventional curvature-driving components. In this approach, heterogeneity is incorporated through a source term on the right-hand side of the LS transport equation, which is defined based on the kinetic properties of the interfaces. The proposed LS formulation is validated by comparing numerical simulation results of quasi-static triple junction (TJ) migration in both 2D and 3D with existing analytical solutions. The simulations accurately reproduce TJ dihedral angles, migration velocities, and profiles, demonstrating excellent agreement with analytical predictions. Furthermore, a more general study on arbitrary TJ migration highlights the method’s broad applicability and strong potential to capture realistic TJ behaviors in real-world scenarios.

Ongoing work focuses on extending the current framework to describe the evolution of multiple TJs in 2D and quadruple junctions in 3D. This extension will enable the simulation of realistic microstructure evolution in polycrystalline materials, with future validation against experimental observations. A key challenge in this regard is the accurate definition of the auxiliary parameter field $\Lambda_i$, which encodes all heterogeneity information. While this study assumes a constant energy per interface, future work will explore methods to incorporate spatially continuous variations in grain boundary (GB) properties. Knowing that in certain polycrystalline systems, GB kinetic properties exhibit complex dependencies and may vary continuously along GBs, e.g., as described by the 5-parameter GB energy model \cite{bulatov_2014}. Additionally, the recently developed disconnection-mediated theory of GB kinetics \cite{han_2018} may be considered to introduce microscopic-scale insights into the LS framework.

\section{Methods}\label{sec4}

The numerical simulations presented in this paper are performed using \textit{Cimlib} \cite{Digonnet_2007,Mesri_2009}, an in-house finite-element library developed at the host laboratory. For 2D simulations, we employ an unstructured triangular mesh with a stabilized P1 solver \cite{bernacki_2011}. To enhance computational efficiency, an adaptive remeshing strategy is adopted \cite{bernacki_2009}. The mesh is isotropically refined within a layer of thickness \( r=2 \times 10^{-2} \) centered at the interfaces, where a uniform mesh size of \( \Delta x_{\text{min}}=1 \times 10^{-3} \) is imposed. Outside this refined layer, a coarser uniform mesh with \( \Delta x_{\text{max}}=2 \times 10^{-2} \) is used. This strategy significantly reduces computational costs while maintaining high resolution for accurately capturing the evolution of triple junctions and their constituent interfaces. For 3D simulations, no adaptive remeshing is performed due to its high computational cost. Instead, we utilize an unstructured tetrahedral mesh with a uniform element size of \( \Delta x=1 \times 10^{-2} \). Time integration is carried out using an implicit Euler scheme, with timesteps set to \( \Delta t=1 \times 10^{-5} \) for 2D simulations and \( \Delta t=2 \times 10^{-5} \) for 3D simulations to ensure numerical stability and accuracy. In accordance with the corresponding benchmark testing configurations, classical von Neumann and Dirichlet boundary conditions are applied in simulations presented in subsections \ref{sec2-2} and \ref{sec2-3}, respectively.

\section*{Acknowledgements}

We thank Carnot M.I.N.E.S and ANR for their financial support (Grant No. 230000488).

\section*{Author contribution}

T.L. developed the model, conducted the simulations, performed the analysis, and wrote the manuscript. M.B. developed the numerical framework, secured the funding and supervised the work. All authors reviewed and approved the final manuscript. 

\section*{Data availability}

The data that support the findings of this study are available from the authors upon reasonable request.

\bibliography{bibliography}

\end{document}